\documentstyle[12pt,epsfig,axodraw]{article}

\begin{document}

\newcommand{\ccaption}[2]{
    \begin{center}
    \parbox{0.85\textwidth}{
      \caption[#1]{\small{{#2}}}
      }
    \end{center}
    }

\thispagestyle{empty}
\vspace*{2cm}
  \begin{Large}\begin{center}
{\sf \sf Double Top Production at Hadronic Colliders }
\end{center}
\end{Large}

\normalsize
\vspace*{0.5cm}
\begin{center}
Yu.P.~Gouz and S.R.~Slabospitsky\\
{\it Institute for High Energy Physics, \\ Protvino, Moscow Region,
142284 Russia}\\
\verb+ gouz@mx.ihep.su+\\
\verb+ slabospitsky@mx.ihep.su+\\
\end{center}                  
\nopagebreak
\vfill
\vskip 3cm
\begin{abstract} 
We calculate the contribution of anomalous top-quark interaction with a gluon,
photon, and $Z$-boson (FCNC) to $t \bar t$ pair production in hadronic
collisions. Using the current data from Run~1 of the Tevatron we evaluate the
upper limit on the the anomalous coupling of $t$-quark with gluon,
$|\kappa_g/\Lambda| \le 0.52$~TeV$^{-1}$. We examine the production of double 
top ($t t$ or $\bar t \bar t$) at hadronic colliders as a result of FCNC
interactions of the top quark. It is shown that the study of such reactions at
LHC collider makes possible to obtain the constraint on strength of coupling at
the level of $|\kappa_g /\Lambda| \le 0.09$~TeV$^{-1}$.
\end{abstract}                                                          

\vskip 1cm
\today \hfill 
\vfill 

\newpage
\setcounter{page}{1}

\section{\bf Introduction }
   
After the discovery of the top-quark at FNAL collider~\cite{disc}, 
the new direction appeared in search for new physics beyond the Standard 
Model~(SM). As the most massive quark in the SM, the $t$-quark is naturally
considered to be more sensitive to new physics than lighter fermions.
Therefore, precision measurements of the top quark properties 
and its production
mechanisms provide unique possibility to obtain information on new physics.

There are numerous speculations on possible new physics manifestations in the
top quark sector. Of  special interest is the study of top quark anomalous
interactions due to flavour--changing neutral currents (FCNC) 
(see~\cite{parke,han95,han96,dirt,yuan,osy,arb,han98}).

Indeed, the absence of elementary FCNC vertices in SM requires the 
consideration of loop contribution, which may generate the effective 
$t \to c (u)$ transitions~\cite{sm}. As a result, the top quark decays via FCNC
are suppressed at the level of $10^{-10}$ to $10^{-12}$, which makes
their observation absolutely impossible.

Various extensions of SM lead to huge enhancement of such transitions.
For example, one can compare the branching ratios of FCNC decays of the
top quarks evaluating in the SM~\cite{sm} and in MSSM~\cite{mssm}:
\begin{eqnarray*}
\begin{array}{l c c c }
 & & {\rm SM} & {\rm MSSM} \\
 {\rm Br}(t \, \to \, c g)      & \approx & 10^{-10} & 10^{-6}\\
 {\rm Br}(t \, \to \, c \gamma) & \approx & 10^{-12} & 10^{-8} \\
 {\rm Br}(t \, \to \, c Z)      & \approx & 10^{-12} & 10^{-8}
\end{array} 
\end{eqnarray*}
Therefore, any observation of FCNC decays would be an indication of a new
physics beyond SM. 

The strategy of searching for anomalous top quark couplings consists of
three complementary approaches: (i) obtain indirect bounds from low energy
processes in which top quark anomalous couplings can enter via
loop processes, or from the bound on $t \rightarrow bW$ which
puts limits on other decay modes of the top, (ii) direct searches
at high energies for the effects of the anomalous couplings in top quark
rare decays, and (iii) the study of the processes of top quark production
via non--SM mechanisms.

The search for rare top decays are extensively considered (see, for
example,~\cite{han95,han96}). The experimental search for these decays
was done by CDF collaboration~\cite{cdf1}. 
One of the possibility for study of the direct top production is considered 
in~\cite{dirt}.  In this scenario, a up (or charm) quark and a gluon from the
colliding hadrons combine immediately to form an s--channel top quark,
namely $u(c) \, + \, g \, \to \, t$. The consideration of other anomalous
processes of the top production can be found in~\cite{yuan,han98}.

In this note we examine a manifestation of anomalous (FCNC)top interaction 
in the process of top--antitop production:
\begin{eqnarray}
p(\bar p) \; p \; \to \; t \; \bar t \; X, \label{ttbar} 
\end{eqnarray}
as well as in the process of double top (or antitop) production:
\begin{eqnarray}
p(\bar p) \; p \; \to \; t \; t \; X. \label{tt} 
\end{eqnarray}

The article is organised as follows. The anomalous FCNC interaction of $t$
quark is considered in Section~2. The available constraints on anomalous 
couplings are considered in Section~3. In Section~4 we evaluate the upper limit
on FCNC top interaction with a gluon from Run~1 of the Tevatron. The estimate
for possible constraint on the same coupling, which may be reached at LHC
collider, is given in Section~5. Section~6 presents general characteristics of
the process of double ($tt$ or $\bar t \bar t$) top production in hadronic
collisions. 
The study of this reaction and possible background 
processes 
are presented in Section~7. The main results are summarized in Conclusion. 

\section {\bf Anomalous $\bar t c$ Vertices }

Following~\cite{peccei90}, the vertices of the FCNC transitions $t \to g c$,
$t \to \gamma c$, and $t \to Z c$ can be written as follows:
\begin{eqnarray}
t g c & \Rightarrow & g_s \frac{\kappa_{g}}{\Lambda}
 \bar t \sigma_{\mu \nu}
[ g_L P_L  + g_R P_R ] t^a c \; G^{a \mu \nu},
 \label{ver1} \\
t \gamma c  & \Rightarrow & e \frac{\kappa_{\gamma}}{\Lambda}
\bar t \sigma_{\mu \nu} 
[ \gamma_L P_L + \gamma_R P_R ] c F^{\mu \nu}, \label{ver2} \\
 t Z c & \Rightarrow & \kappa_z \frac{e}{\sin 2\vartheta_W}
\bar t \gamma_{\mu} [z_L P_L + z_R P_R ] c Z^{\mu}, \label{ver3}
\end{eqnarray}
where $\Lambda$ is the new physics cutoff,
 $\vartheta_W$ is the Weinberg angle, 
$P_{L,R} = \frac{1}{2} (1 \mp \gamma^5)$, 
$\sigma^{\mu \nu} = \frac{1}{2}(\gamma^{\mu} \gamma^{\nu} - \gamma^{\nu} 
\gamma^{\mu})$, 
$\kappa_{g}$, $\kappa_{\gamma}$, and $\kappa_z$ define the strength of the 
anomalous couplings for the currents with a gluon ($\kappa_g$), photon 
($\kappa_{\gamma}$), and $Z$ boson ($\kappa_z$), respectively. The 
magnitudes of left and right components of the currents are denoted by 
$g_{L,R}$, $\gamma_{L,R}$, $z_{L,R}$. They obey the obvious 
constraints: 
\begin{eqnarray*}
 g_L^2 + g_R^2 = 1, \quad  \gamma_L^2 + \gamma_R^2 = 1, \quad
z_L^2 + z_R^2 = 1. 
\end{eqnarray*}
We also assume that ${\rm Im}\,\kappa_g = {\rm Im}\,\kappa_{\gamma} =
 {\rm Im}\,\kappa_z = {\rm Im}\,\gamma_i = {\rm Im}\,g_i = {\rm Im}\,z_i = 0$, 
see~\cite{peccei90,han95}.

Generally speaking, to avoid unitarity violation, the anomalous couplings 
$\kappa$ should 
depend on 
momentum transfer $q^2$. For example,
this dependence may have a form-factor-like behaviour 
\begin{eqnarray} 
\frac{\kappa^2_{g, \gamma}}{\Lambda^2} \Rightarrow 
\frac{\kappa^2_{g, \gamma}}{\Lambda^2 ( 1 + \frac{q^2}{\Lambda^2})}, \quad
 \kappa^2_z  \Rightarrow 
\frac{\kappa^2_z}{1 + \frac{q^2}{\Lambda^2}}. \label{formf}
\end{eqnarray}
However, such a dependence should be essential at very high values of momentum
transfer ($\geq 1$~TeV), and we ignore such dependence in the present
article.

Using the expressions for the vertices~(\ref{ver1}--\ref{ver3}) we find
the equations for the corresponding widths (see also~\cite{han95,dirt,osy}): 
\begin{eqnarray}
\Gamma( t \to c g) &=& \frac{4}{3} \alpha_s m_t^3 
\Biggl ( \frac{\kappa^2_g}{\Lambda^2} \Biggr ), \label{gamglu} \\
\Gamma( t \to c \gamma) &=& \alpha  m_t^3 
\Biggl (\frac{\kappa^2_{\gamma}}{\Lambda^2} \Biggr ), \label{gamgam} \\
\Gamma( t \to c Z) &=&  
\frac{\alpha  \kappa^2_{z} }{8 \sin^2 2\vartheta_W \, M^2_Z} 
m_t^3 \Biggl ( 1 - \frac{M_Z^2}{m_t^2} \Biggr )^2 
\Biggl ( 1 + 2\frac{M_Z^2}{m_t^2} \Biggr ),
\label{gamz}
\end{eqnarray}
where $\alpha$ and $\alpha_s$ are the QED and QCD coupling constants,
respectively, and $M_Z$ is the mass of $Z$ boson.

In what follows we set the masses of the light quark ($c$ or $u$) to zero, 
$m_c = m_u = 0$, and 
\[ 
m_t = 175 \, \; {\rm GeV}. 
\]
We also assume that the anomalous couplings of top quark with $c$ and $u$
quarks are equal in magnitude.

The corresponding values of the widths and branching ratios for different 
decay channels are given in the Table~1.

\section {Current Constraints  on  $|\kappa_g|$ and $|\kappa_{\gamma}|$ }

The analysis of low energy processes (such as  
 $K_L \to \mu^+ \mu^-$, $K_L - K_S$ mass difference, $B^0 - \bar B^0$
mixing, $B \to l^+ l^-X$, $b \to s \gamma$, etc. (see, for 
example~\cite{han95,han96} and references therein)
results in the following constraints:
\begin{eqnarray*} 
\begin{array}{lcl} 
|\kappa_g/\Lambda| &<& 0.95 \;\; {\rm TeV}^{-1}, \\
|\kappa_{\gamma}/\Lambda| &<& 0.28 \;\; {\rm TeV}^{-1}, \\
|\kappa_z| &<& 0.29.
\end{array}
\end{eqnarray*}

CDF collaboration performs the search for decays $t \to \gamma c(u)$ and
$t \to Z c(u)$ at FNAL collider in the reaction of 
$ \bar p p \; \to \; \bar t t X$ at the energy of $\sqrt{s} = 1.8$~TeV.
They obtained the upper limits on branching fractions for the
decays $t \to \gamma c(u)$ and $t \to Z c(u)$~\cite{cdf1} as follows:
\begin{eqnarray}
 {\rm Br}(t \to c \gamma) + {\rm Br}(t \to u \gamma) < 3.2\% 
\quad (95\% \;\;{\rm CL}), \label{dec1} \\
 {\rm Br}(t \to c Z) + {\rm Br}(t \to u Z) < 33\% 
\quad (95\% \;\;{\rm CL}). \label{dec2}
\end{eqnarray}

Using the equations~(\ref{gamgam}) and (\ref{gamz}), one can easily evaluate 
the upper limits on the constants of $\kappa_{\gamma}$ and $\kappa_z$ 
(at $m_t$ = 175~GeV) from the experimental
constraints~(\ref{dec1}) and (\ref{dec2}): 
\begin{eqnarray}
\begin{array}{lcl}
 |\kappa_{\gamma}/\Lambda|  &<& 0.77 \; \; {\rm TeV}^{-1}, 
\label{const1} \\
 |\kappa_z| &<& 0.74. \label{const2} 
\end{array}
\end{eqnarray}

Additional experimental constraints may be obtained from the study of the 
reaction
$e^+ e^-$ annihilation~\cite{osy}
\begin{eqnarray}
e^+ \; e^- \; \to \; \gamma^{\ast} (Z^{\ast}) \; \to \; t \; \bar c \label{ee}
\end{eqnarray}
Starting from Summer 1997, the $e^+ e^-$ collider LEP--2 operates at the energy of
$\sqrt{s} = 184$~GeV~(1997) and 189~GeV~(1998). At this 
in the reaction~(\ref{ee}) becomes kinematically possible,
 and therefore it will be possible to improve significantly  the upper limits 
on the anomalous couplings of top interaction (see~\cite{osy} for details).

\section{Constraint on $|\kappa_g|$ from Run~1 of Tevatron }

In this section we consider what limit on anomalous coupling $|\kappa_g|$
can be derived from the available data on the cross section for 
$t \bar t$ pair production at Run~1 of the Tevatron.  

In order 
to obtain a constraint on $\kappa_g$, we analyse the contribution of anomalous
top interaction into $t \bar t$ production at the FNAL collider. 

Indeed,  both CDF and D$\emptyset$ collaborations measured the top quarks 
production in $b W$ final states. For the case, when any non--SM top 
interaction are very small, one has:
\[
{\rm Br}(t \to W b) \approx 1 \quad {\rm and} \quad 
 \sigma^{\rm exp}_{t \bar t} = \sigma^{\rm SM}_{t \bar t}.
\]
However, for large values of the anomalous couplings the branching ratio to
$Wb$ final state may be substantially smaller than 100\% (see Table~1).
At the same time, there appeared new subprocesses leading to additional
sources of the $t \bar t$ production. In this  case the observed cross
section (in the $b \bar b W^+ W^-$ final state) should be equal to:
\begin{eqnarray}
\sigma (b W^+ \bar b W^-) = {\rm Br}^2(t \to bW) \times 
\sigma ({\rm SM + New}\;\;{\rm Physics}).
\end{eqnarray} 

Therefore, the difference between observed cross section for $t \bar t$ pair
production and that calculated within (SM+New~Physics) framework may be used 
to estimate upper limits on strength of anomalous coupling.

To obtain the corresponding constraints on the anomalous couplings we 
require that
\begin{eqnarray}
| \sigma^{\rm exp}_{t \bar t} - \sigma (b W^+ \bar b W^-)| \le
 2 \Delta, \label{neqn1}
\end{eqnarray} 
where $\Delta$ is an error consisting of both experimental and theoretical
uncertainties.

In the lowest order of perturbative QCD the $t \bar t$ production proceeds via
quark--antiquark and gluon--gluon  annihilation (see~Fig.1a and~1c). 
The anomalous top interactions lead to the appearance of new processes 
in which $t \bar t$ can be produced.
The lowest order diagrams corresponding to these
processes are  presented in Fig.1b~and~1d.

In further analysis we assume that other non--standard decay channels of the 
top  are negligible.

The production cross section measured by the CDF 
collaboration with an integrated luminosity of 110 pb$^{-1}$ is 
$\sigma=7.6{+1.8\atop-1.5}$ pb for $m_t=175$ GeV~\cite{cdf2}, 
which combines dilepton, lepton $+$ jets and all-hadronic channels.  The 
D$\emptyset$  
collaboration gives  $\sigma=5.9\pm1.7$ pb for $m_t=172$ 
GeV~\cite{d02}.  In our analysis we use the  combined result~\cite{sigtop} of 
\begin{equation}
\sigma_{t\bar t}^{\rm exp}=6.7\pm1.3\;\rm pb. \label{combsig}
\end{equation}
For the theoretical cross section in the SM, we adopt the most complete  
result currently available~\cite{qcdtt}, which includes soft-gluon 
summation up to the next-to-leading logarithmic order:
\begin{equation}
\textstyle \sigma^{\rm SM}_{t\bar t}= 5.06{+0.13\atop-0.36}\,\rm pb
\end{equation}
for $m_t=175$ GeV at $\sqrt s=1.8$ TeV.

To make a comparison of the measured ($\sigma^{\rm exp}_{t \bar t}$)
and calculated cross section ($\sigma(b W^+ \bar b W^-)$), 
one should take into account the present uncertainty of the top quark 
mass~\cite{sigtop}, which affects both the experimental and 
theoretical cross sections.  Shifting $m_t$ by $\pm5$ GeV changes  
the theoretical cross section by about $\pm15\%$.  The measured 
cross section also changes with $m_t$ in the same direction, but the 
dependence is weaker.  

Combining all these uncertainties, the possible new physics contribution 
to the cross section is found to be
\[
\Delta = \sqrt{\Delta\sigma^2_{\rm exp} + \Delta \sigma^2_{\rm th}} 
 \approx 1.67\;\;{\rm pb},
 \label{san4}
\] 
where we set~\cite{cdf2,d02} $ \Delta \sigma_{\rm exp} = 1.3$~pb 
(see~(\ref{combsig})), and~\cite{qcdtt}
\[
 \Delta \sigma_{\rm th} = (0.15 \div 0.20) \; \sigma^{\rm SM}_{t \bar t}.
\]

In our case the inequality (\ref{neqn1}) depends essentially on two 
parameters, $\kappa_g$
and $\kappa_z$. The contribution due to a photon exchange 
($\sim \kappa_{\gamma}$) is very small. 

Bearing in mind different notations and normalizations used in the
literature, we present the evaluated constraints on anomalous constants in
terms of constraints on corresponding branching ratios for top quark decays.

The resulting constraints on the branching fractions Br$(t \to g (c+u))$ and 
Br$(t \to Z (c+u))$ are shown in Fig.2. 
The dashed band in this figure
 corresponds to two choices of Br$(t \to \gamma (c+u)) = 0\%$ and 
Br$(t \to \gamma (c+u)) = 3.2 \%$.
The maximum allowed values of these branching ratios
are the following:
\begin{eqnarray}
 {\rm Br}(t \to g (c+u)) & \le & 20\;\% \;\;\; ({\rm at}\;\;\;
{\rm Br}(t \to Z (c+u)) = 0), \\
 {\rm Br}(t \to Z (c+u)) & \le & 32 \;\% \;\;\; ({\rm at}\;\;\;
{\rm Br}(t \to g (c+u)) = 0).
\end{eqnarray}

These constraints correspond to the maximum 
allowed value of $\kappa_g$ (at $\kappa_z \to 0$) as follows: 
\begin{eqnarray}
 |\kappa_g / \Lambda| \leq 0.52 \;\; {\rm TeV}^{-1}.  \label{san5}
\end{eqnarray}

We conclude that the variation of $\sigma_{t \bar t}$ within given uncertainty 
$2\Delta$ does not allow to improve the upper limit on branching fraction
to $Z$--boson channel. On the other hand, from Run~1 data we obtain
the  lowest current constraint on FCNC top quark interaction with a gluon.

The differential distributions of top quarks produced in the reaction
\[
 \bar p \; p \; \to \; t \; \bar t \; X
\]
at $\sqrt{s}$ = 1.8 TeV are shown in Fig.3. It is seen from these pictures
that at the FNAL collider energies the FCNC interaction gives
practically the same form of the differential distributions of the top quarks
like QCD calculations.

\section {\bf Possible Constraints on $|\kappa_g|$ and $|\kappa_z|$ from
$t \bar t$ Production at LHC }

Contrary to the case of the FNAL collider, the contribution of FCNC top quark
interaction leads to significant modification of differential spectra at LHC
energies. Fig.4 presents the corresponding distributions of $t$ quarks
produced in the reaction
\[
 p \; p \; \to \; t \; \bar t \; X
\]
at $\sqrt{s}$ = 14 TeV. In fact, the FCNC interaction results in significant
increase of $t \bar t$ production with high invariant mass 
$M_{t \bar t}$ (see Fig.4). Therefore, the observation of such
deviation from QCD prediction could be considered as a signal from anomalous
top quark interaction\footnote{To be more precise, the shape of these 
distributions is
determined by the point-like behavior of the effective couplings $\kappa$ 
(see explanation of the equation~(\ref{formf})). However, for very high
value of $\Lambda$ (say, $\Lambda \geq$~few~TeV) we can ignore 
$q^2$--dependence of $\kappa$, and our calculated distributions remain
the same.}.

At the same time, the value of this effect depends on the magnitudes of the 
anomalous
couplings, $\kappa_g$, $\kappa_z$, and $\kappa_{\gamma}$. Indeed, for
anomalous coupling values about $\sim 10^{-2} \div 10^{-3}$ this effect
becomes very small. Therefore, the reaction of $t \bar t$ production 
may be not a good method for search for FCNC top quark 
interaction,

Nevertheless, we estimate the possible limits on the anomalous coupling at 
LHC by using the 
same inequality~(\ref{neqn1}). For the difference of
the cross sections calculated in SM and SM+FCNC frameworks (the $\Delta$
parameter) we use: 
\begin{eqnarray}
\Delta_{\rm LHC} = \sqrt{0.15^2_{th} + 0.15^2_{exp}} \times
 \sigma^{\rm SM}_{t \bar t} \approx 175~{\rm pb}. \label{limlhc}
\end{eqnarray}

The corresponding constraint on Br($t \to g (c+u))$ (versus 
Br($t \to Z (c+u))$) is shown in Fig.5. 
The maximum allowed value of this branching ratio is about
\begin{eqnarray} 
 {\rm Br}(t \to g (c+u)) \leq 27 \%. \label{san51}
\end{eqnarray} 
The corresponding maximum allowed value of $\kappa_g$ (at $\kappa_z \to 0$) 
is:
\begin{eqnarray}
 | \kappa_g / \Lambda|  \leq 0.64 \;\; {\rm TeV}^{-1}.  \label{san52}
\end{eqnarray}
Note that this constraint on strength of coupling with a gluon is practically
the same as that evaluated from the data of Run~1 of the Tevatron. Naturally,
 the obtained values of the upper limits~(\ref{san51}) and~(\ref{san52})
are determined  by our assumption on $\Delta_{\rm LHC}$ form~(\ref{limlhc}).

\section {\bf Double Top  Production at Hadronic Colliders }

The FCNC processes result in appearance of the new  processes of 
the top quarks production:
\begin{eqnarray*}
g \;\; u(c) \; & \to & \; t \\
g \;\; u(c) \; & \to & \; t \; Z(\gamma, g) \\
q \;\; \bar q \; & \to & \; t \; \bar c(\bar u) \\
g \;\; g \; & \to & \; t \; \bar c(\bar u) \\
\cdots
\end{eqnarray*}
These are the processes of direct top production, considered, for example,
in \cite{dirt,yuan}. 
  Note, however, that all these processes have huge QCD--background from 
$W + n(jet)$ production, as well as from $t \bar t$ production.

In this article we examine the process which {\bf has no} background from
usual $t \bar t$ production. Namely, we investigate the process of double top
(or antitop) production:
\begin{eqnarray}
 p \; \; p \; \to \; t \, t \; (\bar t \, \bar t) \; \; X. \label{san6}
\end{eqnarray}

Many processes may result in inclusive production of $tt$ ($\bar t \bar t$)
pair in hadron--hadron collisions. For example:
\begin{eqnarray*}
 & \bullet \quad & g g \to t \bar t t \bar t \\
 & \bullet \quad & t t \to t  t \\
 & \bullet \quad & d d \to t W^- t W^- \\
 & \bullet \quad & u \bar d  \to g^{\ast} W^{\ast} \to t
 \bar t t \bar b \\
 & {\rm etc} &
\end{eqnarray*}
However,  most of these reactions are highly suppressed by  higher orders of 
strong and/or electro-weak coupling constants. Other suppressions result from 
 specific kinematics of the production processes.  The non-top final
particles  should be produced in 'invisible' region for detector, i.e. they
should escape  into very forward/backward direction.
 
On the other hand, the $tt$ $(\bar t \bar t )$ pair from the 
reaction~(\ref{san6})
may be produced in a simple $2 \to 2$ subprocess with FCNC interaction.
The corresponding diagrams describing $t t$ production via anomalous top quark
interaction are  shown in~Fig.6.

Naturally, to distinguish this process from conventional $t \bar t$~pair
production one should require that both $t$-quarks  decay semileptonically.
In other words, we consider the reaction with two like-sign isolated leptons
in the final state.

We notice two advantages of the reaction (\ref{san6}):
\begin{itemize}
\item[$\bullet$] it has no background from usual $t \bar t$ production
(because $t \bar t$ pair decays into $l^+ l^-$ pair)
\item[$\bullet$] to detect such a reaction one can apply the same methods as
for the case of $t \bar t$ production.
\end{itemize}

The total cross sections for the reaction~(\ref{san6}) at $\sqrt{s}$ = 14~TeV,
$m_t = 175$~GeV, and $\kappa_g = \kappa_{\gamma} = \kappa_z = 1$ 
are equal to:
\begin{eqnarray*}
 \sigma(t t) &\approx& 480 \; \; {\rm pb}, \\
 \sigma(\bar t \bar t) &\approx& 16 \; \; {\rm pb}.
\end{eqnarray*}
Note that for this choice of the anomalous coupling the relative contributions
due to gluon, photon, and $Z$-boson exchange are
\[
 g \; : \; \gamma \; : \;  Z \; \; \approx \; 70 \; : \; 5 \; : \; 20.
\]
These values of the cross sections correspond to the following number of
events (for the total integrated luminosity of $10^5$~pb${}^{-1}$ and 100\%
efficiency of the isolated lepton detection):
\begin{eqnarray*}
 {\rm N}(l^+ l^+ jet\;jet)_{l=\mu, e} &\approx& 1.9 \cdot 10^6, \\
 {\rm N}(l^- l^- jet\;jet)_{l=\mu, e} &\approx& 6.4 \cdot 10^5. \\
\end{eqnarray*}

Requiring the observation of at least one event with $l^+ l^+ + 2$--jets in
the final state, we can estimate the value of the anomalous coupling
$\kappa_g$ which could be reached at LHC (at $\kappa_{\gamma} = 
\kappa_z = 0$): 
\begin{eqnarray*}
 |\kappa_g / \Lambda| \leq 2.7 \cdot 10^{-2} \; \; {\rm TeV}^{-1}.
\end{eqnarray*}

This value is comparable with that obtained from the study of the reaction of
the direct top production in the subprocess of $g u \to c$ (see \cite{dirt}).

The differential distributions for the $t$ quark from the 
reaction~(\ref{san6}) are shown in the Fig.7. Like for the case of $t \bar t$
production, the FCNC interaction results in significant
increasing of $t t$ production with a high invariant mass of
$M_{t t}$ (see Fig.7).

\section{\bf Signal and Background Calculations }

Previous estimate was obtained in pure 'ideal' case (i.e. 100\% efficiency
for detection of leptons and jets and we assume no background process).
Now we proceed to more 'realistic' estimates.

The major sources of background to the $tt$ production are
\begin{eqnarray}
 q \; q' \; & \to & \; t \; \bar t \; W^{\pm} \label{bcg1} \\
 q \; q \;  & \to & \; W^{\pm} \; q' \; W^{\pm} \; q' \label{bcg2}
\end{eqnarray}

The typical diagrams for each background process are shown in Fig.8 (see
Fig.8a for the  process~(\ref{bcg1}) and Fig.8b for the process~(\ref{bcg2}),
respectively).

The total cross sections for these background processes are (at 
$\sqrt{s}$ = 14~TeV)
\begin{eqnarray}
\begin{array}{l c c l c c}
 \sigma(W^+t \bar t) &=& 0.5 \;{\rm pb}, & 
  \sigma(W^-   t \bar t) &=& 0.24 \; 
  {\rm pb}, \\ 
\sigma(W^+ W^+ qq) &=& 0.5 \;{\rm pb}, & \sigma(W^- W^- qq ) &=& 0.23 \; 
{\rm pb}.
\end{array}
\end{eqnarray}

In order to simulate the detector effects in identifying the signal,
we made a series of standard cuts on the transverse momentum (energy)
$p_T^{}(E_T)$,
pseudorapidity $\eta$, and the jet--lepton separation $\Delta R$. We
call them the 'standard' cuts:
\begin{itemize}
\item[$\bullet$] $p_{\top}^{l} > 15$~GeV,  $|\eta_l| < 2.5$ 
\item[$\bullet$] $\Delta R_{lj} > 0.4$
\item[$\bullet$] $E_{\top}^{jet} > 20$~ GeV, $|\eta_j| < 2.5$
\item[$\bullet$] two isolated like-sign leptons 
\end{itemize}
We take a standard b-tagging with the efficiencies for
$b$--quarks, charmed quarks, and light partons ($u, d, s$ quarks and gluon)
as follows:
\[
\epsilon_{b} = 36\%, \;\; \epsilon_{c} = 10\%, \;\;
\epsilon_{u,d,s} = 1\%.
\]

We use the PYTHIA~5.7~+~JETSET~7.4 Monte Carlo with addition of the routines
calculating the signal and background 
processes. The result of the application of 'standard' cuts is shown
in the Table~2. For the sake of comparison, we started from equal number of 
events ($N_{start} = 10^4$) 
for each reaction, (\ref{san6}), (\ref{bcg1}), and~(\ref{bcg2}).

The resulting  distributions are shown in Fig.9--12.
From these figures one may deduce the difference in the signal and background.

First, in the signal the average number of jets is smaller than in 
background from $t \bar t W$ (see~Fig.9). Both signal and background 
have jets with $E_{\top}^{jet}$ greater than 40--60~GeV (see~Fig.10).
The invariant mass of the isolated charged lepton and jets from 'true'
combinations (i.e. those 
coming from the decay of the same top quark)
can not exceed the top
mass (compare Figs. 11a and 11b). The invariant mass of the two isolated
charged leptons and two jets from the signal is much higher  than 
that for background events (see~Fig.12).

In order to further enhance the signal with respect to background we impose 
the following additional cuts: 

\begin{tabular}{ll}
``two jets'' & we require exactly  two jets with 
``$E^{jet}_{\top} > 20$~GeV; \\ 
``$E_{\top}$'' & the energy of these two jets should be greater than \\
  & 40~GeV, and at least one jet should be tagged as a $b$-jet; \\
``$M(l\;j)$'' & at least in one combination of the lepton+jet pairs 
both \\ & pairs should have invariant masses smaller than 160~GeV \\
& (i.e. $M(l_1 j_1)<160$~GeV and $M(l_2 j_2)<160$~GeV);  \\
``$M(l\;l\;j\;j)$'' & the invariant mass of two charged leptons and 
  two jets \\ & should be greater than 500~GeV. 
\end{tabular}

The result of application of all these additional cuts (step-by-step) are 
shown in the Table~2.  One may deduce that applying all these cuts provide
almost complete  suppression of the background (even at high luminosity
option).

\section{Discussion and Summary}

We may use the results of the signal and background calculations to
determine the maximum allowed value of $\kappa_g/\Lambda$ 
 that could be reached at LHC collider. We use the following criterion 
\[
 N_S \geq 3 \sqrt{N_S+N_B},
\]
where $N_S$ and $N_B$ are the number of signal and
background events, respectively. This criterion corresponds 
 approximately  to the $95 \%$ confidence level.

For the total luminosity of 
$\int {\cal L} dt = 100$~fb$^{-1}$ we have $N_B = 2.7$ (for 
$\mu \mu$ mode) and $N_B = 10.8$ (for $ll = \mu \mu + \mu e + ee$) 
(see Table~2).
As a result we obtain (for $\kappa_z = \kappa_{\gamma} = 0$)
\begin{eqnarray}
|\kappa_g / \Lambda|  & \leq & 0.116_{\mu \mu}
(0.091_{ll}) \;\;\;{\rm TeV}^{-1}, \\
{\rm Br}(t \to g \; (c+u)) &\leq& 1.2 \times 10^{-2}{}_{\mu \mu}
( 7.4 \times 10^{-3}{}_{ll}).
\end{eqnarray}

In summary, we calculate the contribution for $t \bar t$ pair production
resulted from the anomalous top quark interaction (FCNC). From the analysis of
the data from Run~1 of the Tevatron we deduce the lowest current limit on 
the anomalous coupling $\kappa_g / \Lambda < 0.52 \rm{~TeV}^{-1}$.

We investigate the double  top--quark production at LHC
collider via the chromo--magnetic flavor--changing neutral current
couplings $\bar tcg$ and $\bar tug$. We find that the strength of the
anomalous coupling $\bar t q g$ may be probed to $\kappa_g / \Lambda =
0.09 \rm{~TeV}^{-1}$ at the LHC with $100 \rm{~fb}^{-1}$ of data at 
$\sqrt{s} = 14$~TeV.   Assuming $\kappa_g \equiv 1$, the scale of
new physics $\Lambda$ can be probed to 11~TeV  at the LHC.

\section {\bf Acknowledgements}

We thank to A.~Arbuzov, M.~Cobal, F.~Froidevaux, M.~Mangano, R.~Mekhdiev, 
J.~Parsons, V.~Obraztsov, A.~Zaitsev, and O.~Yushchenko for  fruitful 
discussions. This work was supported, in part, 
by Russian Foundation for Basic Research,
projects no.~96-15-96575.

\newpage

\begin{table}

{\bf Table~1. } The values of the anomalous couplings and corresponding branching
fractions for top quark decay, calculated within SM framework and by using
the values resulting from the Tevatron Run~1 data and from low-energy data.
The mass of the top quark is $m_t = 175$~GeV, $\kappa / \Lambda$ is 
in~TeV$^{-1}$.

\begin{center}
\vspace {0.5cm}

\begin{tabular}{|l|c|c|c|c|c|} \hline 
 & {\rm channel} & $W^+ b$ & $q g$ & $q \gamma$ & $q Z$ \\ \hline
SM &$\kappa / \Lambda$ && $8 \cdot 10^{-6}$ & $3\cdot 10^{-6}$ & 
$\kappa_z \simeq 4 \cdot 10^{-7}$ \\ \hline
& Br  & 1. & $5\cdot 10^{-11}$ & $5\cdot 10^{-13}$ &
$1.5 \cdot 10^{-13}$  \\ \hline \hline
&$\kappa / \Lambda$ && $ 1 $ & $1$ & 
$\kappa_z = 1 $ \\ \hline
 & Br  & $0.35$ & $0.32$ & $0.02(0.05)$ & $0.32$   \\ \hline \hline
FNAL & $\kappa / \Lambda$ && $ 0.5 $ & $0.77 $ & 
$\kappa_z = 0.74 $ \\ \hline
 & Br  & $0.57$ & $0.13$ & $0.018 $ & $0.28$ \\ \hline \hline
Low-energy &$\kappa / \Lambda$ && $ 0.5 $ & $0.28$ & 
$\kappa_z = 0.29 $ \\ \hline
 & Br  & $0.77$ & $0.17$ & $0.003$ & $0.056$  \\ \hline 
\end{tabular}

\end{center}
\end{table}

\vspace{1cm}

\begin{table}

{\bf Table~2. } Result of the application of differential cuts for the
processes of $t t$ production and relevant background reactions 
(see Section~7 for detail explanation).

\begin{center}
\vspace {0.5cm}

\vspace{5mm}

\begin{tabular}{|l|c|c|c|} \hline 
  & signal $t t$ & Bckg. $W t \bar t$ 
& Bckg. $q W q W$ \\ \hline \hline
cross section (for $\kappa = 1$) & 0.67 pb & 0.01 pb & 0.01 pb\\ \hline \hline
 cut & $N_{events}(t t)$ & $N_{events}(W t \bar t)$ 
& $N_{\rm events}(q W q W)$ \\ \hline \hline
 no cut  & $10000$ & $10000$ & $10000$ \\ \hline 
"standard" cuts & $3452$ & $5354$ & $5462$\\ \hline
"two jets" & $2095$ & $712$ 
& $3269$ \\ \hline
"$E_{\top}$" & $1221$ & $316$ & $118$ \\   \hline
$M(l j) < 160$~GeV & $1177$ & $190$ & $46$  \\ \hline
$M(ll j j) > 500$~GeV & $853$ & $15$ & $12$ \\ \hline \hline
$N_{events}$ for $L = 10^5$ pb${}^{-1}$ & $5715$ & $1.5$ & $1.2$ \\ \hline
\end{tabular}
\end{center}
\end{table}

\newpage
\begin{figure}[t]
\begin{center}
\epsfig{file=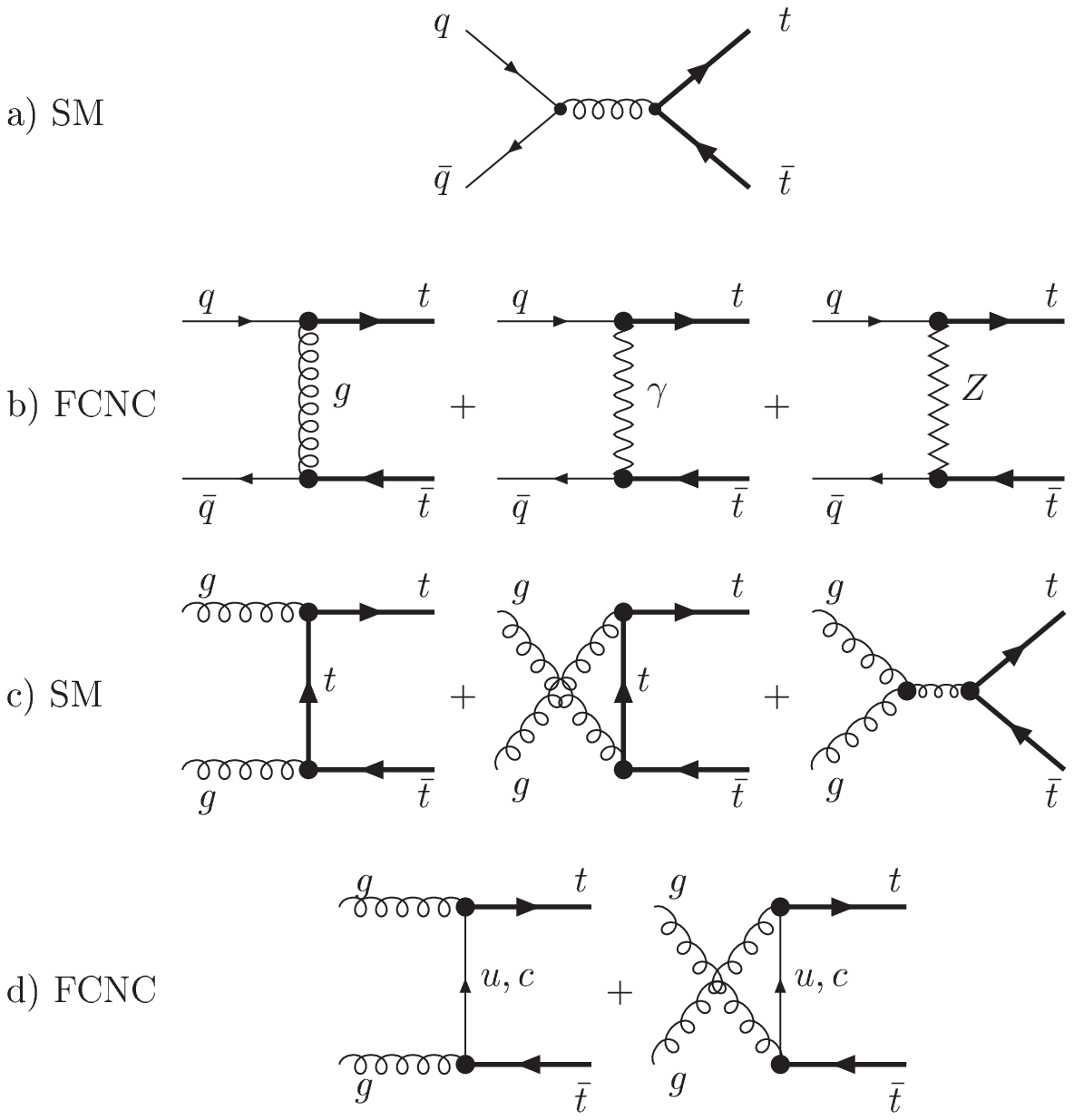,height=16cm,width=12cm,clip=}
\ccaption{}{
Lowest order diagrams describing $t \bar t$ production in QCD (a and c) and
in FCNC (b and d).
}
\end{center}
\end{figure}

\begin{figure}[t]
\begin{center}
\epsfig{file=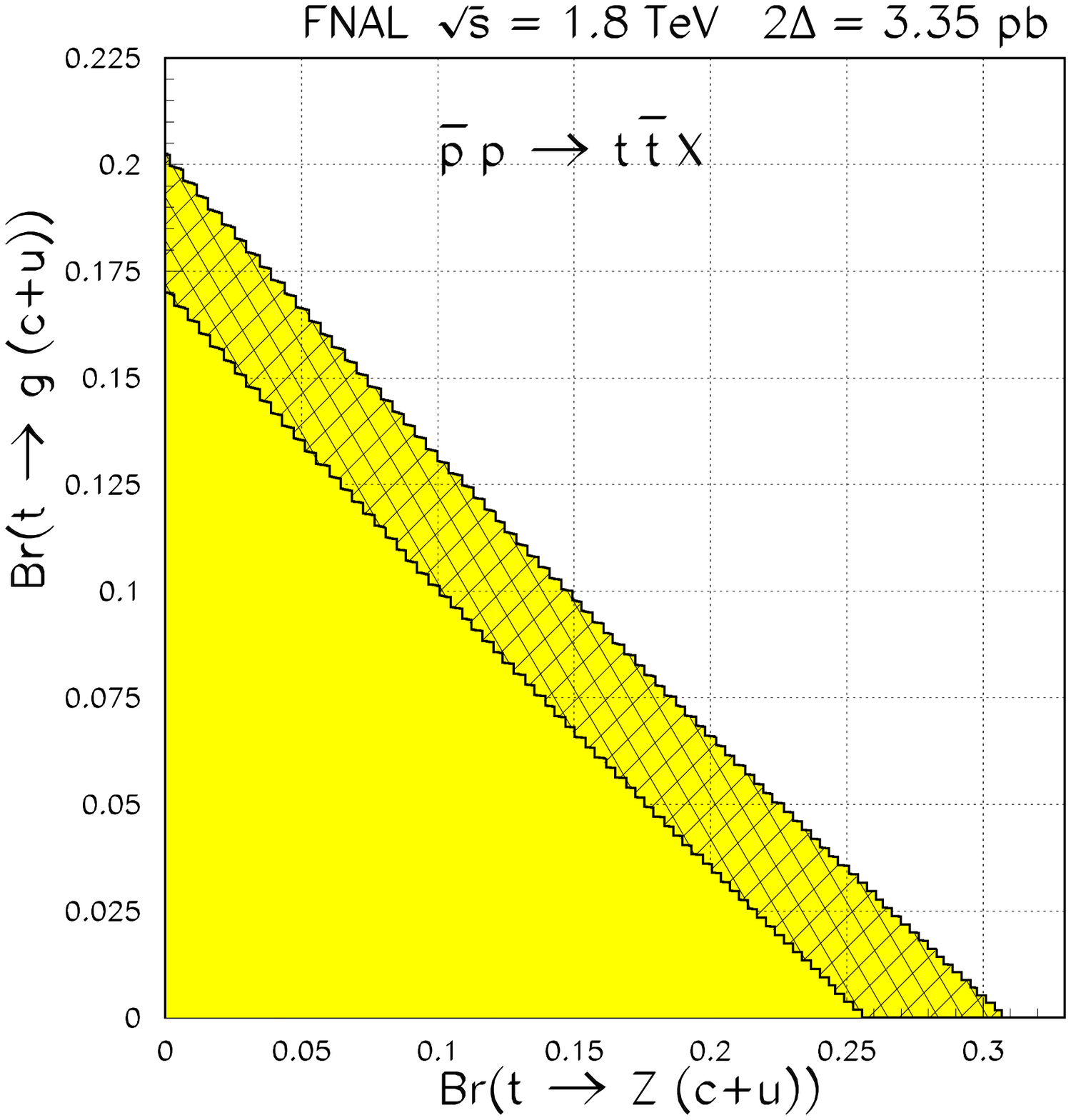,width=12cm,clip=}
\ccaption{}{
Upper limit on Br($t \to g (c+u)$) versus Br($t \to Z (c+u))$ resulted from 
the analysis of the $t \bar t$ production at Run~1 of the Tevatron.
}
\end{center}
\end{figure}

\begin{figure}[t]
\begin{center}
\epsfig{file=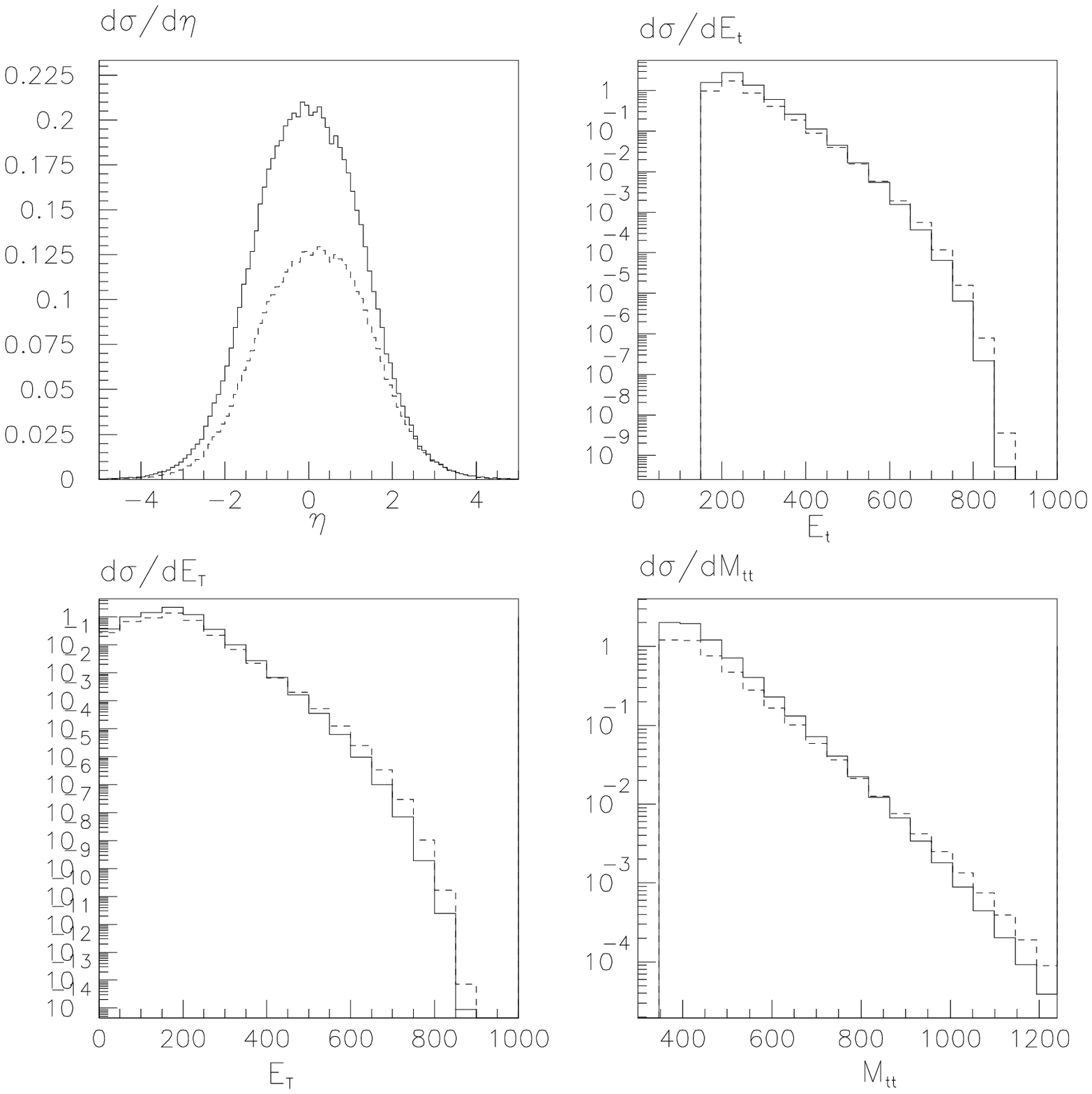,width=12cm,clip=}
\ccaption{}{
Differential distributions over pseudorapidity ($\eta$), 
$t$ quark energy ($E_t$), transverse energy of the top ($E_T$), and invariant
mass of the $t \bar t$ pair ($M_{tt}$), for $t$ quark produced in the reaction
$\bar p p \to t \bar t X$ at $\sqrt{s}$ = 1.8~TeV (FNAL collider). Solid 
curves are QCD predictions, dashed curves are QCD+FCNC
calculations. All distributions are given in arbitrary units. 
}
\end{center}
\end{figure}

\begin{figure}[t]
\begin{center}
\epsfig{file=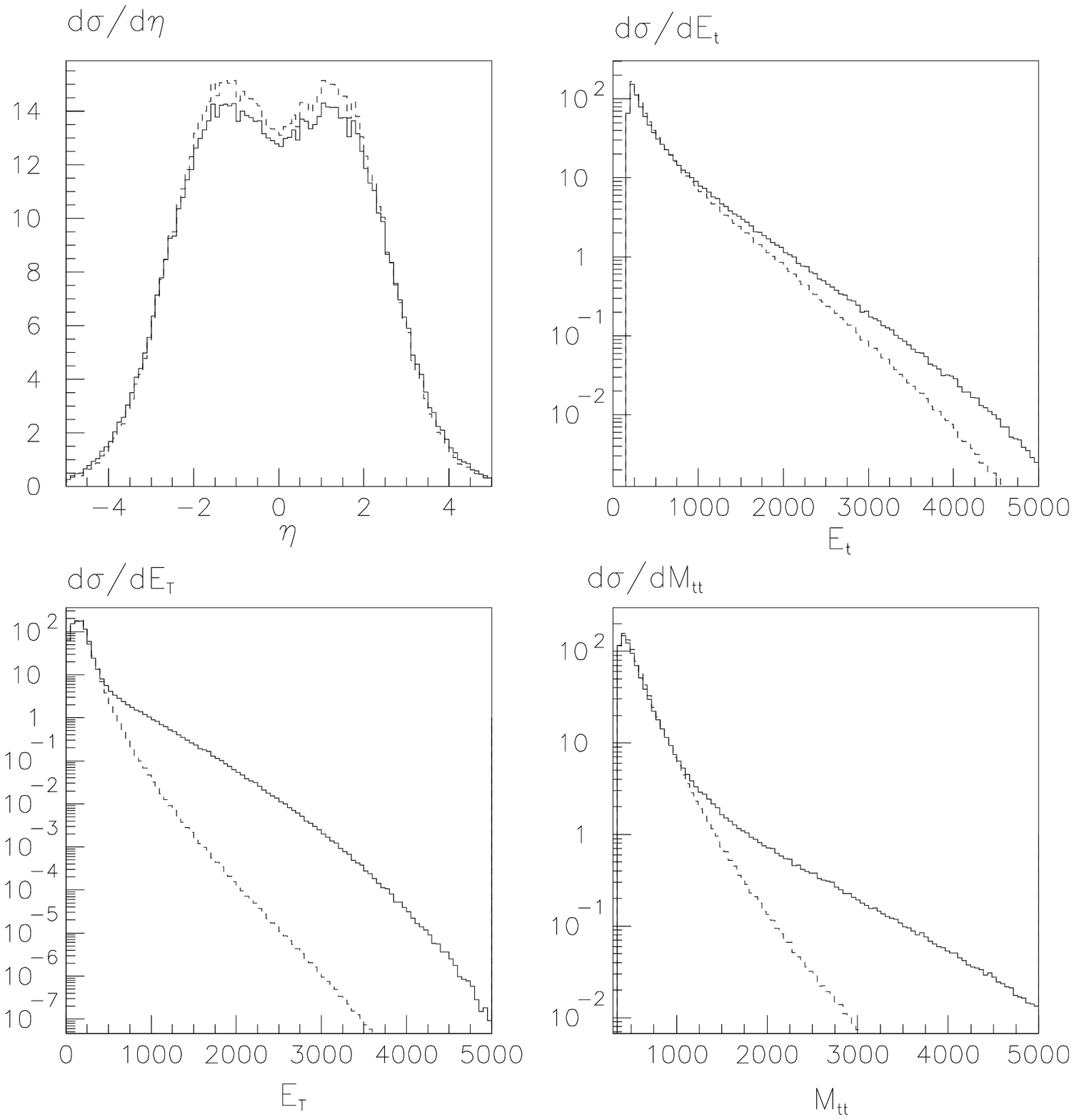,width=12cm,clip=}
\ccaption{}{
The same distributions as in Fig.3, but calculated for the reaction
$p \; p \; \to \; t \; \bar t \; X$ at $\sqrt{s}$ = 14~TeV (LHC collider). 
}
\end{center}
\end{figure}

\begin{figure}[t]
\begin{center}
\epsfig{file=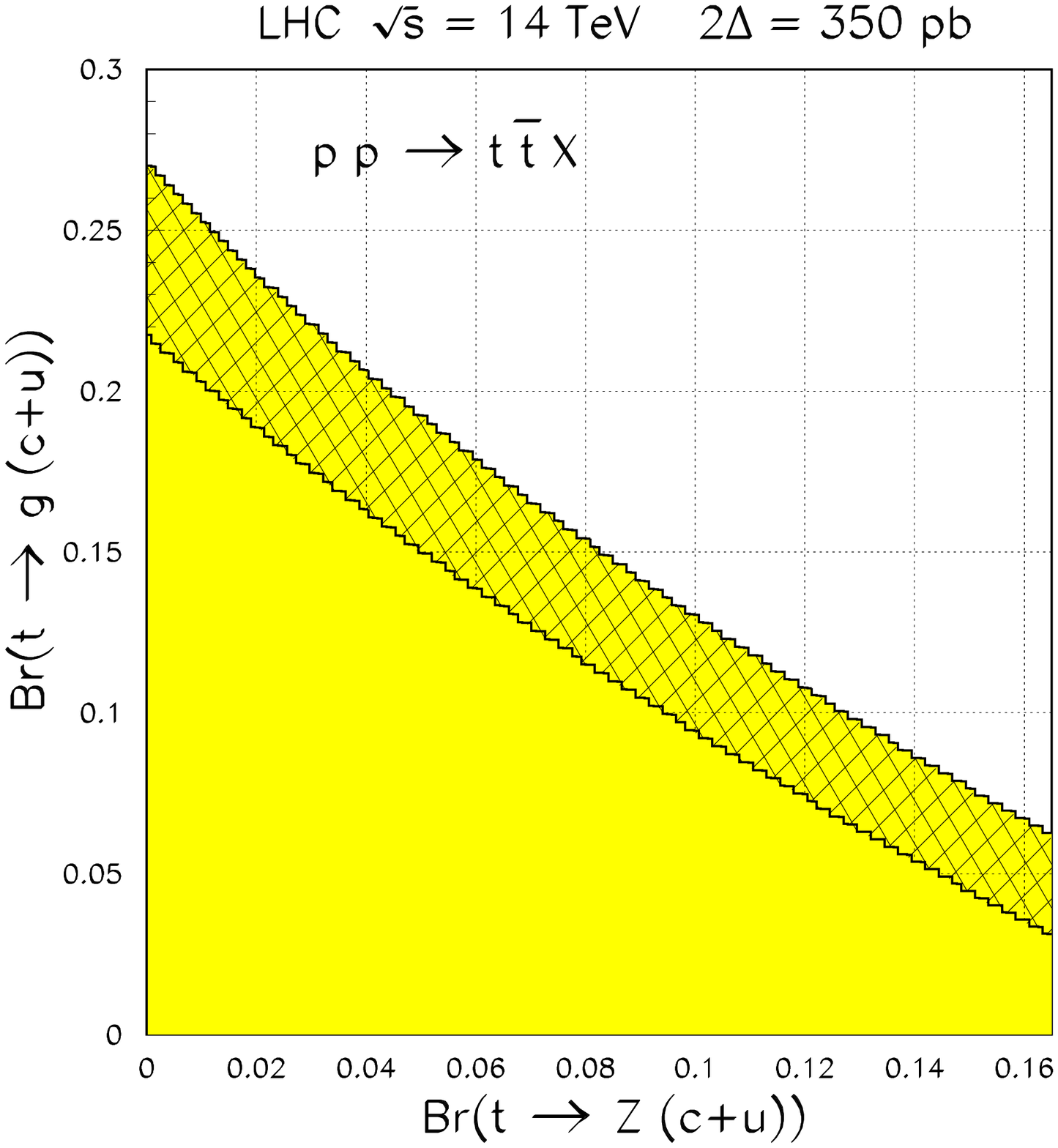,width=12cm,clip=}
\ccaption{}{
Upper limit on Br($t \to g (c+u)$) versus Br($t \to Z (c+u))$ which may be 
obtained at LHC collider from analysis of the $t \bar t$ pair production 
cross section.
}
\end{center}
\end{figure}

\begin{figure}[t]
\begin{center}
\epsfig{file=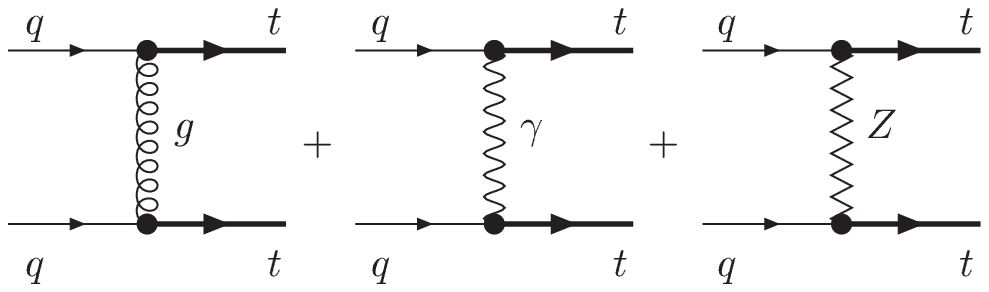,width=12cm,clip=}
\ccaption{}{
Diagrams describing $t t $ quarks production due to FCNC interaction.
}
\end{center}
\end{figure}

\begin{figure}[t]
\begin{center}
\epsfig{file=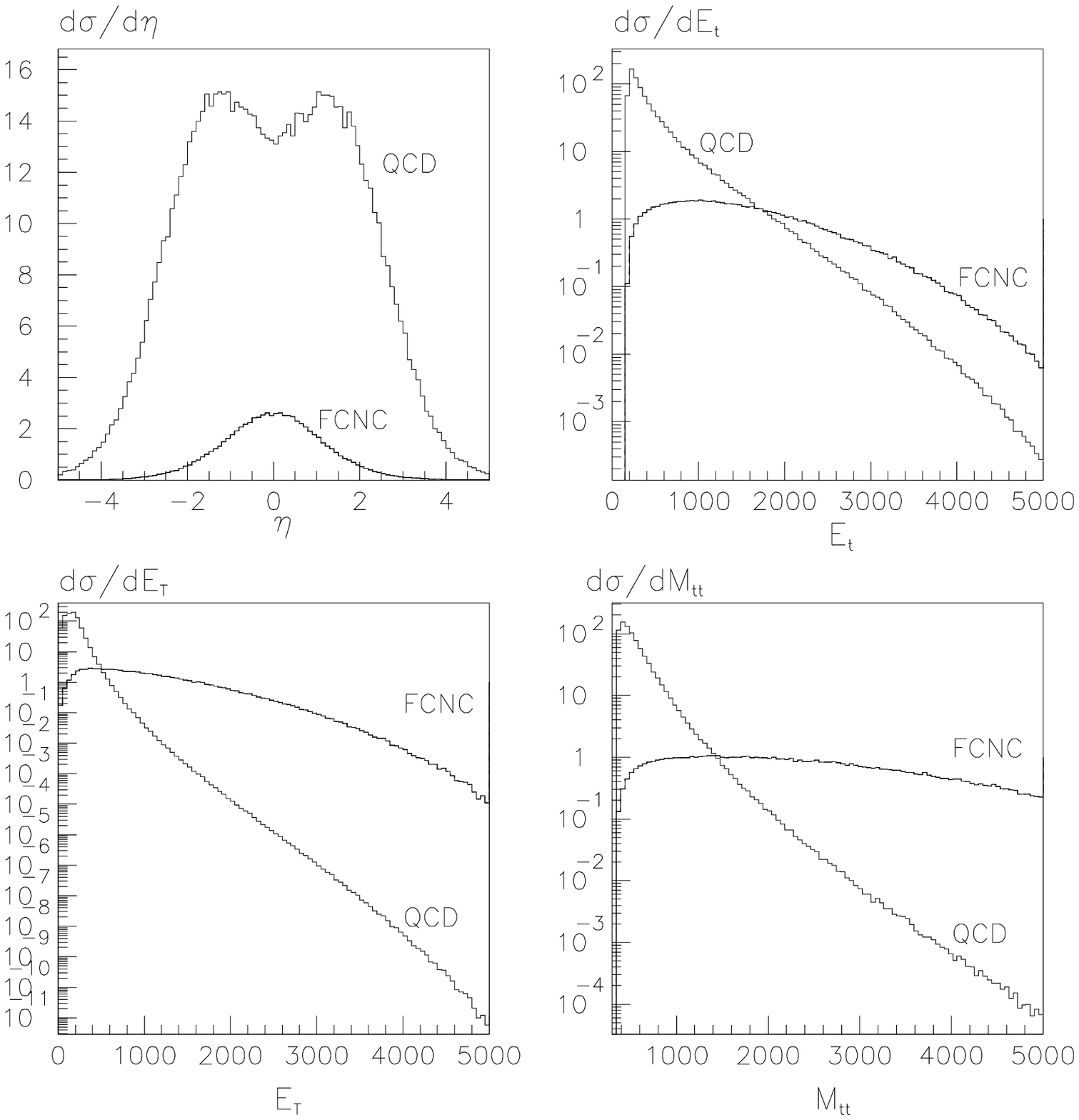,width=12cm,clip=}
\ccaption{}{
Differential distributions on pseudorapidity ($\eta$), 
$t$ quark energy ($E_t$), transverse energy of the top quark ($E_T$), and invariant
mass of the top-top pair ($M_{tt}$), for $t$ quark produced in the reaction
$p p \to t t X$ at $\sqrt{s}$ = 14~TeV (LHC collider). 
All distributions are given in arbitrary units. (For comparison we show at the
same plots the QCD predictions for the case of $t \bar t$ production at the
same energy).
}
\end{center}
\end{figure}

\begin{figure}[t]
\begin{center}
\epsfig{file=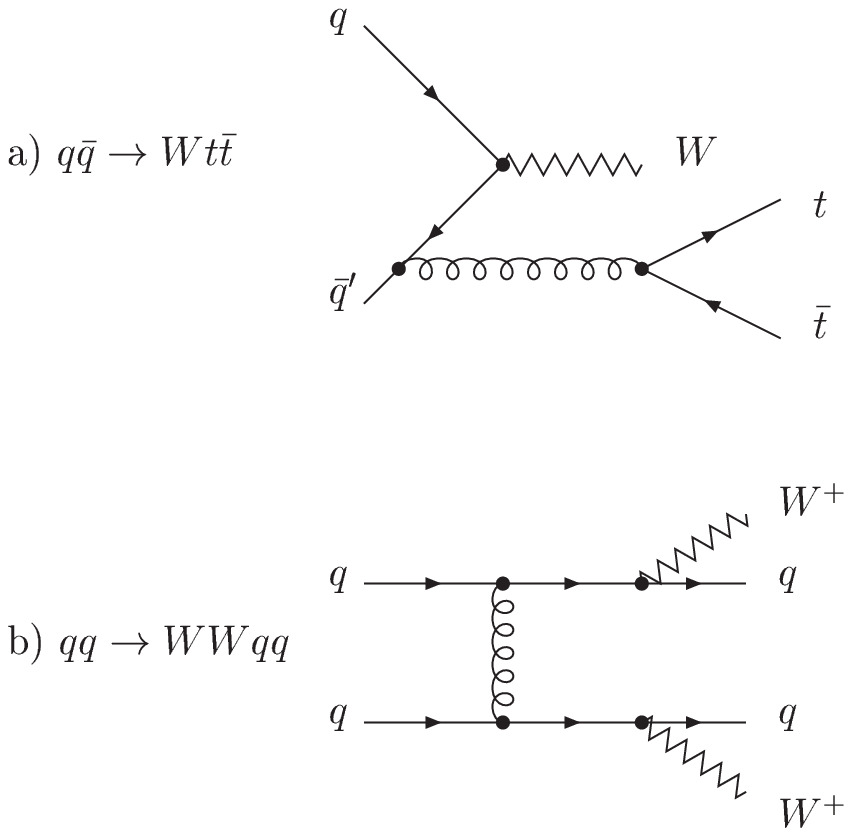,width=12cm,clip=}
\ccaption{}{
Diagrams describing the background processes for the reaction of $t t$ pair
production.
}
\end{center}
\end{figure}

\begin{figure}[t]
\begin{center}
\epsfig{file=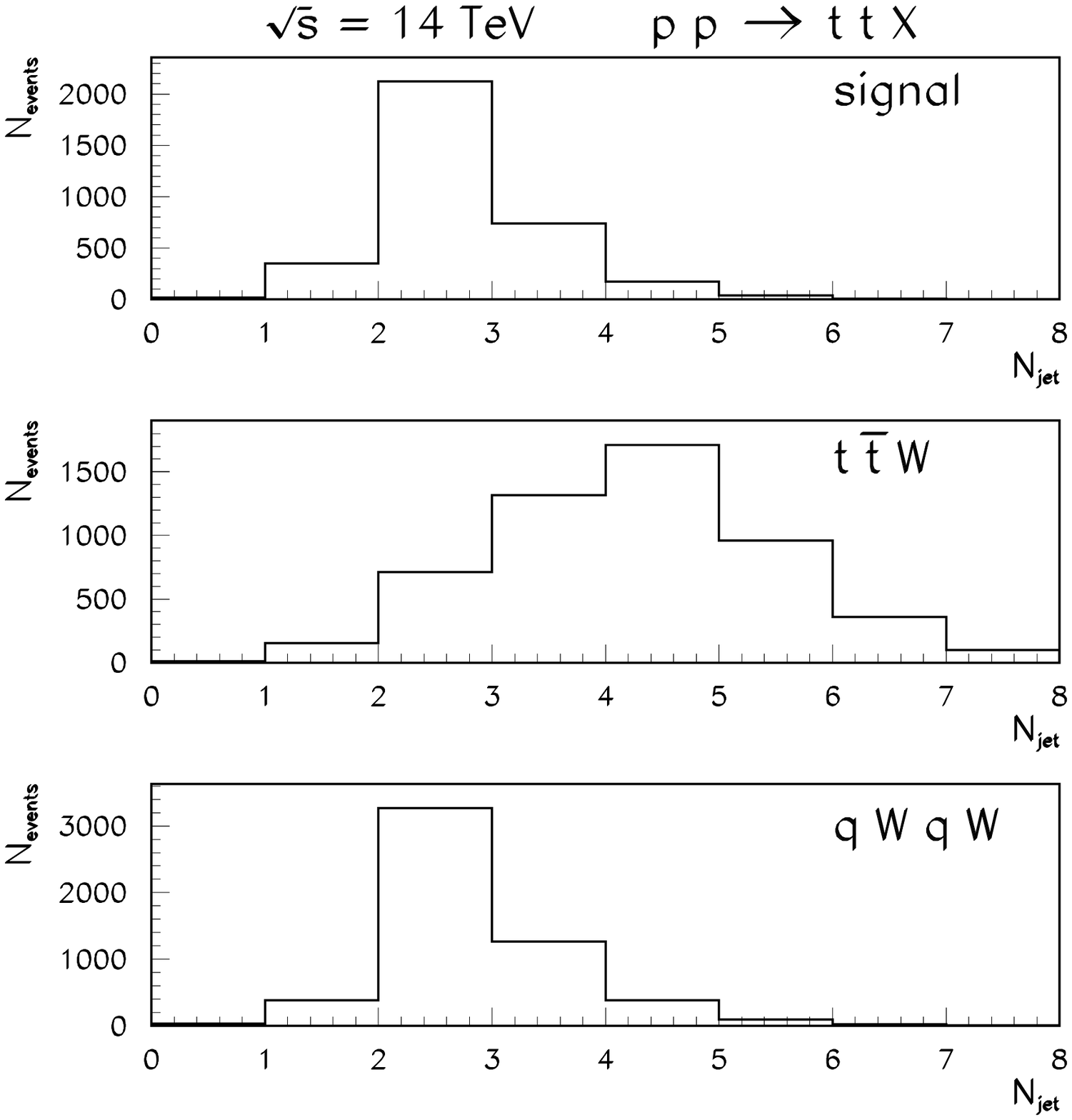,width=12cm,clip=}
\ccaption{}{
 Distributions on number of jets  $N_{jet}$.
}
\end{center}
\end{figure}

\begin{figure}[t]
\begin{center}
\epsfig{file=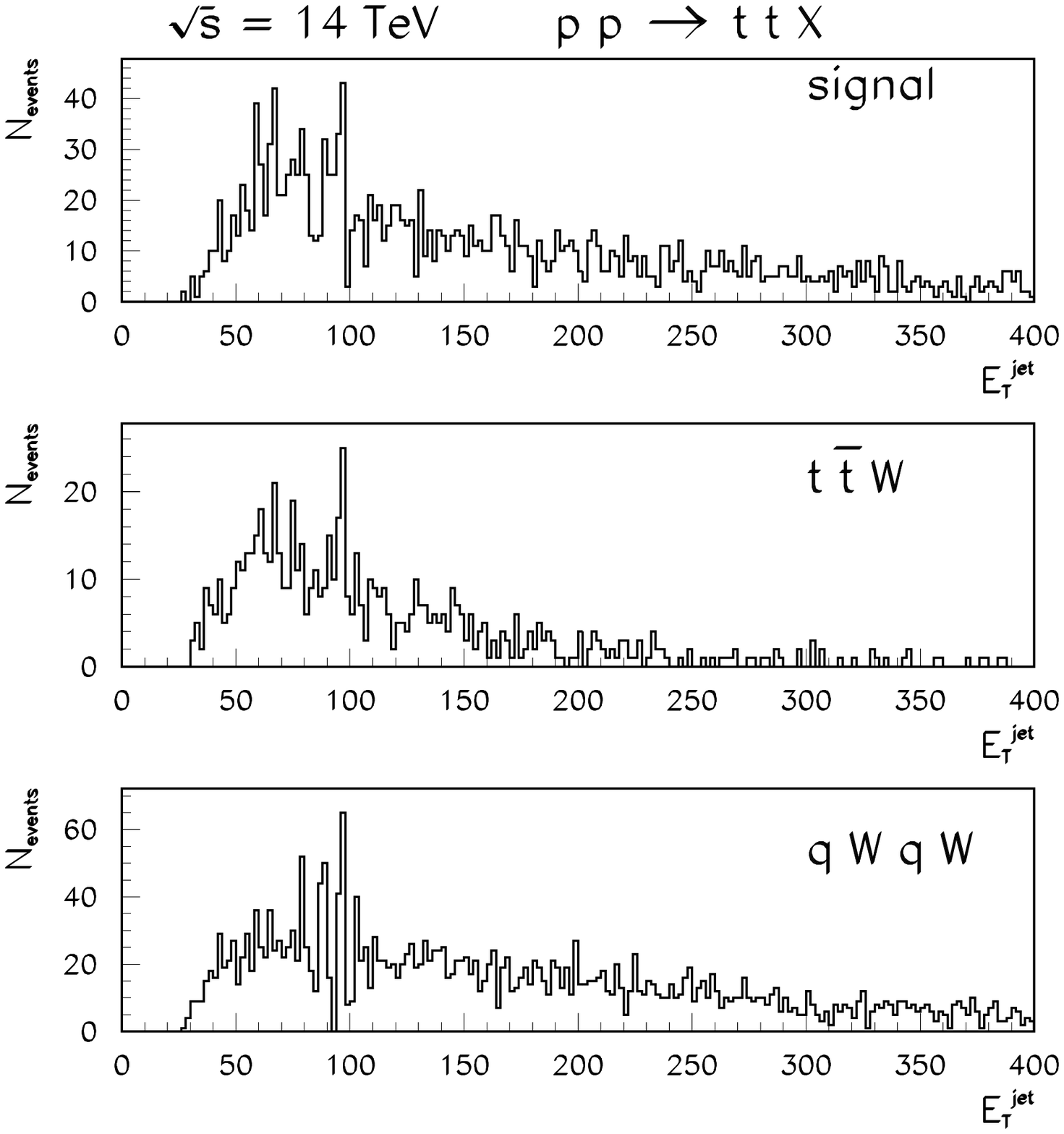,width=12cm,clip=}
\ccaption{}{
 $E_{\top}^{jet}$--distributions from the signal ($t t$) and background
processes. $E_{\top}^{jet}$ is in GeV.
}
\end{center}
\end{figure}

\begin{figure}[t]
\begin{center}
\epsfig{file=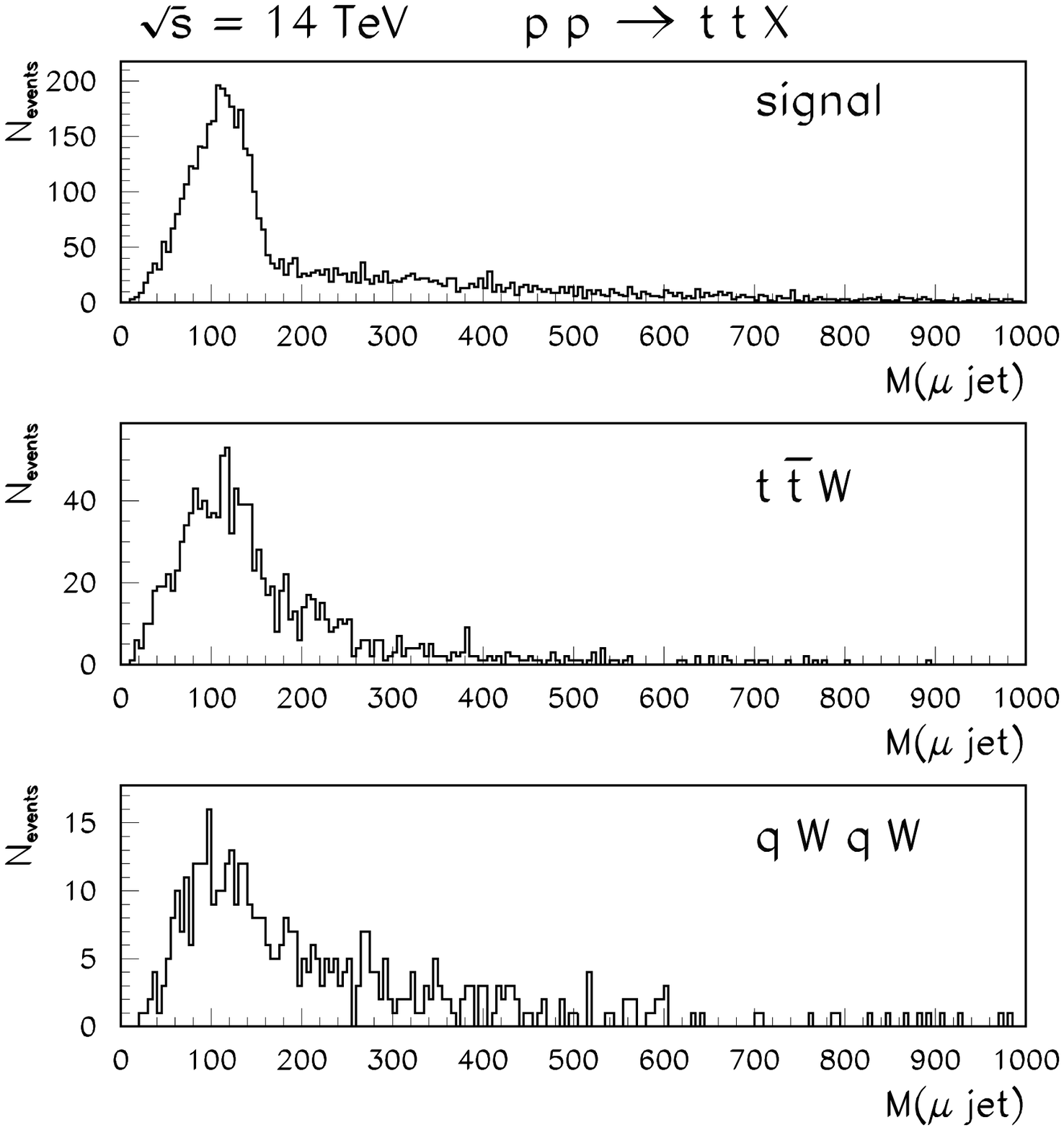,width=12cm,clip=}
\ccaption{}{
Distributions on invariant mass of charged lepton and jet.
All combinations of leptons and jets are taken. $M(\mu\, jet)$ is in GeV.
}
\end{center}
\end{figure}

\begin{figure}[t]
\begin{center}
\epsfig{file=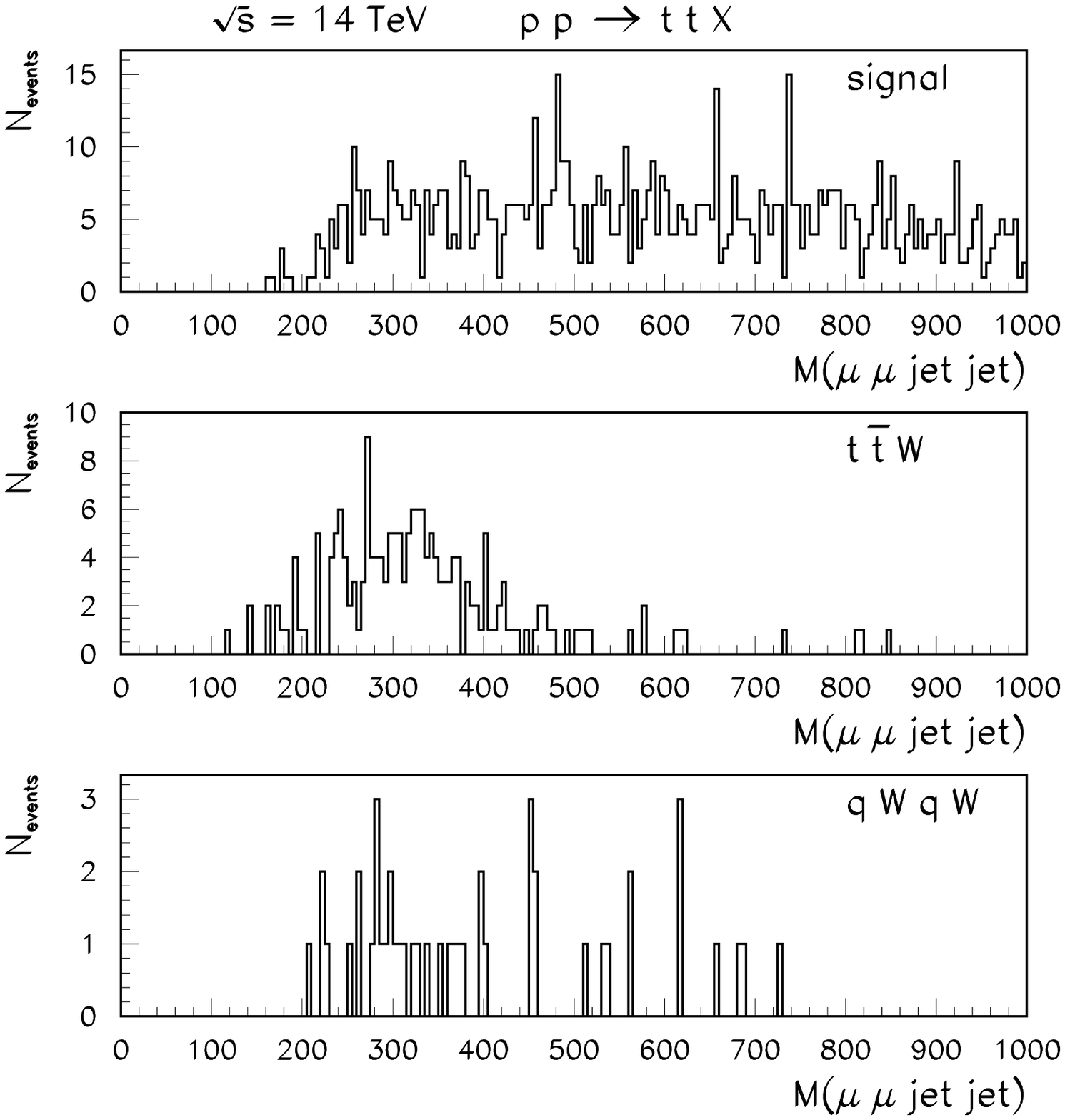,width=12cm,clip=}
\ccaption{}{
Distributions on invariant mass of two charged leptons and 
two jets. $M(\mu\, \mu\, jet\, jet)$ is in GeV.
}
\end{center}
\end{figure}

\end{document}